\documentclass[prl,twocolumn]{revtex4}
\usepackage{epsfig}
\usepackage{bm}
\usepackage{amsmath}

\begin{document}

\title{Distributed chaos and turbulence in B\'{e}nard-Marangoni and Rayleigh-B\'{e}nard convection}

\author{A. Bershadskii}

\affiliation{
ICAR, P.O. Box 31155, Jerusalem 91000, Israel
}

\begin{abstract}
  Temporal and spatio-temporal (turbulence) distributed chaos in B\'{e}nard-Marangoni and Rayleigh-B\'{e}nard convection have been studied using results of laboratory experiments and direct numerical simulations in the terms of effective chaotic diffusivity (viscosity) and action. It is shown that for the both cases decaying part of the power spectra has stretched exponential form - for temporal spectrum $E(f) \propto \exp-(f/f_{\beta})^{1/2}$ and for spatial spectrum $E(k) \propto \exp-(k/k_{\beta})^{2/3}$, where the $f_{\beta}$ and $k_{\beta}$ represent low-frequency (large-scale) coherent oscillations. 
\end{abstract}

\maketitle

\section{Chaotic (turbulent) diffusivity and action}

  The conception of effective turbulent diffusivity have been widely used both in the theory and in the numerical simulations \cite{my} (we include in this term an effective 'renormalization' of molecular viscosity, thermal conductivity and molecular diffusivity in general). Simplest way to estimate the chaotic (turbulent) diffusivity $D$ is
$$
D=v_cl_c   \eqno{(1)}
$$
where $v_c$ and $l_c$ are the characteristic velocity and spatial scale respectively. In the wavenumber terms $l_c \propto 1/k_c$
$$
D \propto v_ck_c^{-1}  \eqno{(2)}
$$
Then
$$
v_c \propto D~k_c  \eqno{(3)}
$$
In the frequency terms this estimate can be rewritten using the dimensional considerations as 
$$
v_c \propto D^{1/2} f_c^{1/2}  \eqno{(4)}
$$
where $f_c$ is characteristic frequency. This estimate resembles the basic relationship of the Hamiltonian systems \cite{suz}
$$
v_c \propto I^{1/2} f_c^{1/2}   \eqno{(5)}
$$
where $I$ is an action. This resemblance has a deep physical meaning for theories based on the Navier-Stokes equations (see, for instance, Ref. \cite{E} and references therein). Actually
$$
D = c I   \eqno{(6)}
$$
where $c$ is a dimensionless constant and $I$ is a chaotic (stochastic \cite{E}) action.

\section{Distributed chaos}  

For smooth dynamical systems with compact strange attractors decaying part of chaotic frequency spectrum is often exponential \cite{fm}-\cite{sig}
$$
E(f) \propto \exp-(f/f_c) \eqno{(7)},
$$    
 If the characteristic frequency $f_c$ is ensemble varying parameter, then in order to find a spectrum one should compute the ensemble average
$$
E(f ) \propto \int_0^{\infty} P(f_c) \exp -(f/f_c)~ df_c  \eqno{(8)}
$$
where $P(f_c)$ is the ensemble probability distribution of the $f_c$. 

   For Gaussian distribution of the characteristic velocity $v_c$ \cite{my} the characteristic frequency $f_c$ can be found from Eq. (4) (or Eq. (5))
$$
P(f_c) \propto f_c^{-1/2} \exp-(f_c/4f_{\beta})  \eqno{(9)}
$$
where $f_{\beta}$ is a constant. 

   Substituting Eq. (9) into Eq. (8) we obtain
$$
E(f) \propto \exp-(f/f_{\beta})^{1/2}  \eqno{(10)}
$$

  Now let us compute spatial (wavenumber) spectrum using spatial relationship Eq. (3) instead of the temporal Eq. (4) (or Eq. (5)). Let us in analogy with the temporal (frequency) stretch exponential spectrum Eq. (10) assume that the spatial (wavenumber) spectrum is also a stretched exponential:
$$
E(k) \propto \int_0^{\infty} P(k_c) \exp -(k/k_c)dk_c  \propto \exp-(k/k_{\beta})^{\beta}  \eqno{(11)}
$$
 Asymptote of $P(k_c)$ at $k_c \rightarrow \infty$ can be estimated from the Eq. (11) as 
$$
P(k_c) \propto k_c^{-1 + \beta/[2(1-\beta)]}~\exp(-bk_c^{\beta/(1-\beta)}) \eqno{(12)}
$$
here $b$ is a constant \cite{jon}. On the other hand, it follows from the Eq. (3) that for Gaussian distribution of the characteristic velocity $v_c$ the characteristic wavenumber $k_c$ also has Gaussian distribution. To make the asymptotic distribution Eq. (12) Gaussian one should take $\beta = 2/3$, i.e.
$$
E(k) \propto \exp-(k/k_{\beta})^{2/3}  \eqno{(13)}
$$
for the spatial (wavenumber) spectrum.

\section{Temporal chaos in the B\'{e}nard-Marangoni convection}

At the B\'{e}nard-Marangoni convection a horizontal fluid layer is heated from a plate below
and cooled from above through a free upper surface. Unlike the Rayleigh-B\'{e}nard convection, where buoyancy is the main physical factor, in the B\'{e}nard-Marangoni convection the free surface tension effects play a significant role as well. Therefore, the Marangoni number - Ma, is added to the the Prandtl and Rayleigh numbers used for description of the Rayleigh-B\'{e}nard convection. \\

  In the Ref. \cite{rca} results of a laboratory experiment with the B\'{e}nard-Marangoni convection in a small
hexagonal vessel (aspect ratio $\Gamma = 2.2$) filled with silicone oil were reported. The lateral walls were thermally insulating. A vertical laser beam, which is reflected from the bottom of the vessel, was used for the measurements (a non-intrusive method). A laser spot was observed on a screen. The displacements of the spot centre were measured in order to obtain a signal. Both the interface deformation and the variation of the refraction index under influence of the thermal gradients result in the observed beam deflections. \\

 With increasing of the Marangoni number the mono-periodic and  bi-periodic states, temporal and then spatio-temporal chaos were successively observed. For the temporal chaos a spatial order is not dynamically changing despite the temporal fluctuations, whereas for the spatio-temporal chaos the interaction between temporal and spatial modes results in a dynamic spatial disorder.\\

  Figure 1 shows (in the semi-log scales) spectrum of the signal for the temporal chaos state: $Ma = 543$; $Pr = 440$ (the spectral data were taken from Fig. 5 of the Ref. \cite{rca}). 
The dashed curve is drawn to indicate the stretched exponential spectrum Eq. (10). The dotted arrow indicates position of the $f_{\beta}$. One can see that the entire temporal distributed chaos is tuned to the low-frequency ($f_{\beta}$) oscillations. 

  The correlation dimension \cite{gp} characterising the strange attractor at $Ma = 543$ and $Pr = 440$ is approximately equal to 6.7. 
 
\begin{figure} \vspace{-1.8cm}\centering
\epsfig{width=.45\textwidth,file=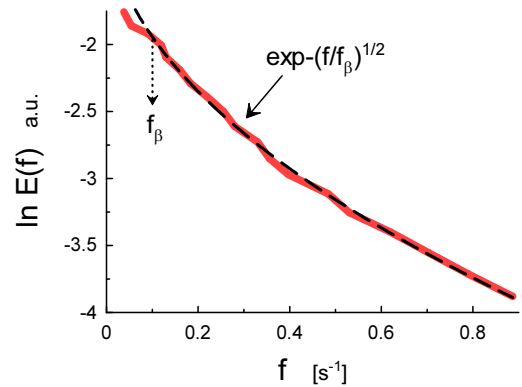} \vspace{-4.2cm}
\caption{Power spectrum of the signal for the temporal chaos state: $Ma = 543$; $Pr = 440$.} 
\end{figure}

\section{Temporal and spatial spectra in the Rayleigh-B\'{e}nard convection}

  Spectrum close to Eq. (10) was observed for the first time in the turbulent Rayleigh-B\'{e}nard convection for temperature temporal fluctuations and reported in Ref. \cite{wu}. The measurements were performed in the centre of an upright cylindrical cell with helium gas. At this configuration, however, one cannot exclude a non-zero mean velocity in the centre of the cell (as a consequence of large-scale circulation, see for instance \cite{sbn} and references therein) and due to the Taylor hypothesis \cite{my} the measured frequency spectrum cannot be considered as pure temporal one \cite{kv}. \\  
\begin{figure} \vspace{-0.5cm}\centering
\epsfig{width=.35 \textwidth,file=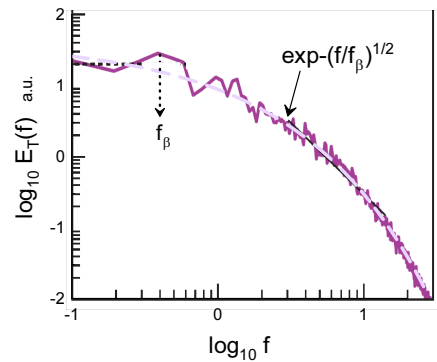} \vspace{-2.5cm}
\caption{Temporal power spectrum for the temperature fluctuations at $Pr =1$ and $Ra=10^8$.} 
\end{figure}
\begin{figure} \vspace{-1cm}\centering
\epsfig{width=.42\textwidth,file=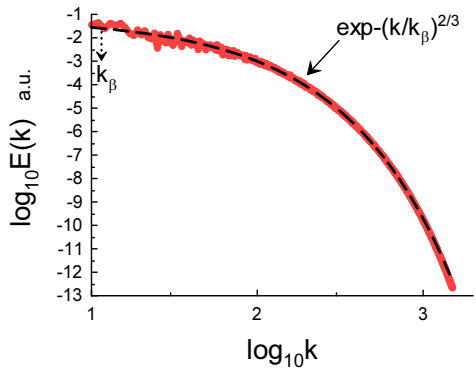} \vspace{-4.1cm}
\caption{Spatial kinetic energy spectrum at $Pr =1$ and $Ra=10^8$. } 
\end{figure}

     Figure 2 shows (in the log-log scales) a frequency power spectrum for the temperature fluctuations in a direct numerical simulations of the Rayleigh-B\'{e}nard convection (the spectral data were taken from Fig. 7 of the the Ref. \cite{kv}). The measurements were performed using real-space probes at centre of a cubical box. The Boussinesq dynamical equations
$$
\frac{\partial{\bf u}}{\partial t}+({\bf u\cdot\nabla){\bf u}}  = -\nabla p+ T {\bf e}_z+ \sqrt{\frac{\mathrm{Pr}}{\mathrm{Ra}}}\nabla{}^{2}{\bf u}  \eqno{(14)},
$$
$$
\frac{\partial T}{\partial t}+({\bf u}\cdot\nabla)T  =  \frac{1}{\sqrt{\mathrm{Ra}\mathrm{Pr}}}\nabla{}^{2}T \eqno{(15)},
$$
$$
\nabla\cdot{\bf u}  =  0  \eqno{(16)},
$$
were numerically solved in this box. For the temperature field $T(t,{\bf x})$ at the horizontal walls (cooled from above and heated from below) conducting boundary conditions were applied, whereas at the side  walls insulating boundary conditions were used. For the velocity field ${\bf u} (t,{\bf x})$ the no-slip boundary conditions were used at all the walls. The Prandtl number $Pr = 1$ and Rayleigh number $Ra = 10^8$ in this DNS. 

   It is important that there was no mean velocity at the centre of the cube and, therefore, the real-space probes' measurements provide purely temporal spectrum \cite{kv}. The blue dashed curve in the Fig. 2 is drawn to indicate the stretched exponential spectrum Eq. (10). The dotted vertical arrow indicates the frequency $f_{\beta}$. The distributed chaos can be considered as tuned to the low-frequency coherent oscillations with the frequency $f_{\beta}$.  \\

 In recent Ref. \cite{vvs} results of a direct numerical simulations of the Rayleigh-B\'{e}nard convection at the same $Ra=10^8$ and $Pr =1$ were reported.  The free-slip boundary conditions for the velocity field and conducting boundary conditions for the temperature field were used at the top and bottom horizontal plates, whereas at the side walls of the computational domain the periodic boundary conditions were used and random initial condition were applied.

  Figure 3 shows spatial kinetic energy spectrum obtained in this DNS. The dashed curve is drawn in order to indicate the stretched exponential spectrum Eq. (13) (cf. Fig. 2) and the dotted arrow indicates $k_{\beta}$.\\

\section{Acknowledgement}

I thank R. Samuel and M.K. Verma for sharing their data, and G.L. Eyink for sending his paper and comments.

\end{document}